\documentclass[journal]{IEEEtran}
\usepackage{blindtext, graphicx}
\usepackage{color}
\usepackage{cite}
\usepackage{amsmath,amssymb}
\usepackage{siunitx}
\usepackage[normalem]{ulem}
\begin{document}
\renewcommand{\arraystretch}{1.15}
%\title{Stability issues of multi-microgrids}
\title{High-Fidelity Model Order Reduction for Microgrids Stability Assessment}

\author{Petr~Vorobev,~Po-Hsu~Huang,~\IEEEmembership{Student Member,~IEEE}, Mohamed~Al~Hosani,~\IEEEmembership{Member,~IEEE}, James~L.~Kirtley,~\IEEEmembership{Life Fellow,~IEEE}, and Konstantin~Turitsyn,~\IEEEmembership{Member,~IEEE}

% \thanks{This work was funded by the Cooperative Agreement between the Masdar Institute of Science and Technology, Abu Dhabi, UAE and the Massachusetts Institute of Technology, Cambridge, MA.}
% \thanks{P. Vorobev and K. Turitsyn are with with Department of Mechanical Engineering,Massachusetts Institute of Technology. \:E-mail: petrvoro turitsyn@mit.edu}
% \thanks{xxxx are with Department of Electrical and Computer Engineering, Massachusetts Institute of Technology. \:Email:jelizon, pohsu, kirtley@mit.edu}
% \thanks{xxxx are with Department of Electrical Engineering and Computer Science, Institute Center for Energy, Masdar Institute of Science and Technology. \:Email:xxx@masdar.ac.ae}
}

% \author{\IEEEauthorblockN{P.Vorobev\IEEEauthorrefmark{1}\IEEEauthorrefmark{2}}

% \IEEEauthorblockA{\IEEEauthorrefmark{1}Department of Mechanical Engineering\\
% Massachusetts Institue of Technology\\
% Cambridge, MA}

% \IEEEauthorblockA{\IEEEauthorrefmark{2}Skolkovo Institute of Science and Technology\\
% Moscow, Russia}
% }

\maketitle

\begin{abstract}
Proper modeling of inverter-based microgrids is crucial for accurate assessment of stability boundaries. It has been recently realized that the stability conditions for such microgrids are significantly different from those known for large-scale power systems. While detailed models are available, they are both computationally expensive and can not provide the insight into the instability mechanisms and factors. In this paper, a computationally efficient and accurate reduced-order model is proposed for modeling the inverter-based microgrids. The main factors affecting microgrid stability are analyzed using the developed reduced-order model and are shown to be unique for the microgrid-based network, which has no  direct analogy to large-scale power systems. Particularly, it has been discovered that the stability limits for the conventional droop-based system ($\omega-P/V-Q$) are determined by the ratio of inverter rating to network capacity, leading to a smaller stability region for microgrids with shorter lines. The theoretical derivation has been provided to verify the above investigation based on both the simplified and generalized network configurations. More importantly, the proposed reduced-order model not only maintains the modeling accuracy but also enhances the computation efficiency. Finally, the results are verified with the detailed model via both frequency and time domain analyses. 
\end{abstract}

\begin{IEEEkeywords}
Droop control, microgrids, reduced-order model, small-signal stability.
\end{IEEEkeywords}

\IEEEpeerreviewmaketitle

%----------------------------------------------------%
%               S E C T I O N  I
%----------------------------------------------------%      

\section{Introduction}\label{sec:introduction}
The advances in the renewable energy harvesting technologies and evergrowing affordability of electrical storage devices naturally lead to increased interest in microgrid development. Microgrids are expected not only to be an effective solution to geographically remote areas, where the interconnection to the main power grid is infeasible, 
but also are considered as an improvement for conventional distribution networks during their disconnection from the feeding substation \cite{hatziargyriou2007microgrids}. While in grid connected mode, the simplest and most commonly used 
method of operation is to set renewable sources to maximum power output with the grid's interconnection taking responsibility for any power imbalances. With the increasing share of distributed generation and, more importantly, in the islanded mode of operation, there is a need for proper control of individual inverters power output \cite{hatziargyriou2007microgrids}. The problem of designing a proper control for microgrids has been the subject of intensive research in the last two decades. Comprehensive reviews \cite{olivares2014trends, Zoka2004proc, Jiayi2008comprehensive, parhizi2015state} on the state-of-the-art in the field give an insight to the main approaches  utilized for microgrid control.    

One of the first propositions for inverters connected to an AC grid were made more than two decades ago \cite{chandorkar1993control} with a droop control based on real power-frequency and reactive power-voltage control loops. These control methods were proposed to replicate conventional schemes utilized at large-scale central power generations for proper load sharing. The stability issue of microgrids operation was first recognized in \cite{coelho2002small} and \cite{guerrero2004wireless} where small-signal stability analyses are carried out in a way similar to transmission grids. 
By looking at the mathematical and physical models utilized in these studies, there was no principle difference between microgrids and transmission grids and, hence, all principles of small-signal stability which are valid for large scale power systems can be applied to microgrids. It was later realized that a high $R/X$ ratio, which is typical for microgrids, can lead to considerable changes in microgrid stability regions \cite{hatziargyriou2013microgrids}. While the analysis and modeling of large-scale power system has been thoroughly investigated in the literature, there is far less experience and systematic studies of microgrids modeling with justification and validation. A natural question is weather the microgrids are similar to large-scale power systems or if there is a qualitative difference between them with certain phenomena specific to microgrids.

Modeling of microgrids, as any other engineering system, relies heavily on the appropriate choice of simplifications. With respect to small-signal stability analysis, the main question is weather a particular model reduction technique can give qualitatively incorrect results (i.e., predicting stability while in reality the system is unstable or vice-versa). A full model for stability assessment of microgrids was developed in \cite{pogaku2007modeling} considering all internal states of the inverter as well as network dynamics. The method was later extended for microgrids with dynamic load responses. 
%While full models are the most reliable in stability assessment, there are two easily become very complex and computationally demanding with the increase in the system size,; b) it is very difficult to get an insight into the key factors influencing stability ;c) they require a lot more modeling efforts on detailing the process, increasing the chance of modeling errors. It is clear that reliable reduced-order models are essential not only to reduce the computational requirements but for attaining a mathematical insight into physical origins of instability. 
While full-order models are the most reliable in stability assessment, they suffer from the following: a) full-order models can easily become very complex and computationally demanding with the increase in the system size; b) it is very difficult to get an insight into the key factors influencing stability; c) they require a lot more modeling efforts which increases the chance of modeling errors. It is clear that reliable reduced-order models are essential not only to reduce the computational requirements but also to attain a mathematical insight into physical origins of instability. 

The first attempts to model microgrids were made following the experience from large-scale power systems while neglecting the network dynamics \cite{coelho2002small,guerrero2004wireless,chandorkar1993control}. This approach seems reasonable since the timescales of network dynamics are determined by electro-magnetic transient times which are very small (in the order of few milliseconds) for resistive microdgids ($X/R$ ratio is around unity). One of the first, to our knowledge, reduced-order models that captured the effects of network dynamics was developed in \cite{iyer2010generalized} where the grid dynamics was incorporated by a certain perturbation method. The importance of network dynamics despite its very fast nature was pointed out in \cite{guo2014dynamic} where a similar perturbation approach was used. In \cite{mariani2015model}, the inadequacy of oversimplified models was further emphasized where it was explicitly shown that in certain situations, the full-order model predicts instability while the reduced-order (Kuramoto's) model predicts stability for a wide range of parameters. A model reduction technique based on singular perturbation theory was introduced in \cite{rasheduzzaman2015reduced} allowing for proper exclusion of fast degrees of freedom, which is based on the formal summation of multiple orders of expansion in powers of small parameters (time scales ratio).

It is indicated in previous literature work that a simple timescale ratio is a poor indicator for the significance of a certain degree of freedom - even very fast network dynamics can influence the slow modes. It is also obvious, that along with accuracy and computational simplicity, the reduced-order models should be transparent enough allowing for analysis of major factors affecting stability. There is, therefore, a need to develop a technique that properly separates fast and slow degrees of freedom which simplifies the calculation procedures without significant reduction in accuracy.

This paper concentrates on developing a systematic approach with special emphasis on the physical mechanisms of fast state variables' participation in slow modes. The key contributions of this paper are as follow:
\begin{enumerate}
    \item A reliable and concise reduced-order model for microgrids stability assessment is developed allowing for fast and accurate stability studies as well as dynamic simulations. 
    
    \item The influence of fast degrees of freedom on system dynamics is properly quantified and the reasons for inadequacy of quasi-stationary (with respect to network dynamics) approximation are given.  
    
    \item Generalization of the proposed method to arbitrary sets of slow and fast degrees of freedom is presented as well as an explicit form of reduced-order equations for microgrids of arbitrary structure.      
\end{enumerate}

%\textcolor{blue}{(deleted) } %The rest of the paper is organized as follows: in Section \ref{sec:2bus}, the problem is formulated on a two-bus model and explicit influence of fast degrees of freedom is illustrated. Also, a physical explanation of instability mechanism is provided and phenomena, specific to microgrids, are discussed. Section \ref{sec:genap} gives a formal mathematical formulation of the problem and presents a general solution. Section \ref{net} describes an application of the mathematical model to microgrids with arbitrary network structure. Section \ref{sec:num} gives an illustrative example of the method on a five inverter-based microgrid and presents explicit comparison of our proposed method to exact modeling and quasi-stationary approximation. 

%----------------------------------------------------%
%               S E C T I O N  II
%----------------------------------------------------%      

\section{Two-Bus Model}\label{sec:2bus}
% Assumptions
% 5-state model
% Stability 3-state and 5-state model: discussion
% Single vs multiple inverters 
% Reduced model
%
\begin{figure}[t!]
\centering
        \includegraphics[width=0.32\textwidth]{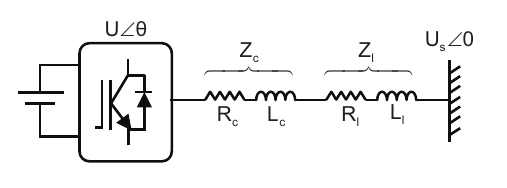}
        \caption{Simplified network configuration of a two-bus model.}
        \label{fig:2bus}
\end{figure}

In this section, the microgrid stability problem that motivates this study is illustrated using a simple two-bus system % example %
shown in Fig. \ref{fig:2bus}. % We consider a% 
The method of high-fidelity order reduction is then  generalized to a network setting in Section \ref{net}.

A single inverter with nominal apparent power $S_n$ and coupling impedance with resistance $R_c$ and inductance $L_c$ is connected to an infinite bus with fixed voltage $U_s\angle0$. The connecting line is characterized by inductance $L_l$ and resistance $R_l$. The inverter operates in a droop mode, such that the equilibrium frequency, voltage and generated powers are related to each other via the standard droop relations:  $\omega = \omega_{set} - m_p P$ and $U= U_{set} - n_q Q$, where $\omega_{set}, U_{set}$ are the inverter terminal frequency and voltage  settings, respectively; and $P,Q$ are the instantaneous active and reactive power generated by an inverter, respectively. The values of $m_p$ and $n_q$ are the frequency and voltage droop gains, respectively. For analysis purposes, it is also convenient to introduce the dimensionless droop gains $k_p = m_p S_n/\omega_0$ and $k_q = n_q S_n/U_0$ that characterize the relative response of the powers to frequency and voltage variations. %We note, that the droop gains $k_p$ and $k_q$ are normalized to the individual inverter rating $S_n$ (which may be different for different inverters in the system) thus representing a natural relative gain of each inverter. Typically, the values of $k_p$ and $k_q$ are set within $0.5\% - 3\%$ \cite{pogaku2007modeling}.
It should be noted that the droop gains $k_p$ and $k_q$ are normalized to the individual inverter rating $S_n$ (which might be different for different inverters in the system) thus representing a natural relative gain of each inverter. Typically, the values of $k_p$ and $k_q$ are set within $0.5\% - 3\%$ \cite{pogaku2007modeling}.

\iffalse

Practically, the droop-based inverters %does not intend to be operated in the grid-tied  
are not designed to operate in grid-connected mode due to high stiffness of the grid. The inverter will eventually reach the frequency of the grid, and any amount of power that satisfies the $\omega-P$ droop equation will have to be supplied by the inverter, which can be unacceptable. Therefore, in this section, the case under study is presented for illustration purposes only.

\fi

Throughout this paper, the following representation for voltage and current will be utilized: 
\begin{equation}\label{VIrep}
v(t)=Re[V(t)e^{j\omega_0 t}];\quad i(t)=Re[I(t)e^{j\omega_0 t}],
\end{equation}
where the complex amplitudes $V(t)$ and $I(t)$ can be arbitrary (not necessarily slowly varying) functions of time. The frequency $\omega_0$ is the equilibrium frequency of the system which for a grid-connected inverter coincide with the grid frequency. The index $0$ is used throughout the paper to denote the equilibrium values of corresponding variables. It should be noted that these equations represent a mathematical change of variable and do not introduce any approximation to the dynamic equations. Similar representation is used in \cite{iyer2010generalized} and \cite{guo2014dynamic}. 

The full-order model of the inverter includes a number of internal degrees of freedom with different timescales - a description of such a model can be found in \cite{pogaku2007modeling}. It is also noted there, as well as in \cite{mohamed2008adaptive} and \cite{nikolakakos2016stability}, that the set of modes associated with power controllers are of main interest from the point of view of stability. While in general one can not omit the fast degrees of freedom, it is legitimate to do with the inverter internal states, leaving a model where only the terminal voltage and frequency are effectively controlled. Therefore, the stability of the two-bus system can be described with sufficient accuracy by the following model with three states related to the inverter and two states for the line. The inverter states are the terminal voltage magnitude $U$, the phase angle $\theta$ (both can be combined to a phasor $V(t)=U(t)e^{j\theta(t)}$) and the frequency $\omega$. The line states are the $dq$ components of the current. The equations for the described model are: 
\begin{align}
    \frac{d \theta}{dt} &= \omega-\omega_0 \label{thetaeq}
\\  \tau\frac{d \omega}{d t} &= \omega_{set}-\omega - m_p P \label{omegaeq}
\\  \tau\frac{d U}{dt} &= U_{set}-U - n_q Q\label{Veq}
\\  L\frac{d I_d}{dt} &= U \cos\theta - U_s - R I_d + \omega_0 L I_q%
\label{ieqD}
\\  L\frac{d I_q}{dt} &= U \sin\theta - RI_q - \omega_0 L I_d%
\label{ieqQ}
\end{align}
Here, equations \eqref{omegaeq} and \eqref{Veq} represent the dynamics of the terminal voltage and frequency, and incorporate the internal %inverter control system filters%
 power filters of the inverter control system characterized by the bandwidth $w_c = \tau^{-1}$. Equations \eqref{ieqD} and \eqref{ieqQ} model the electromagnetic dynamics of the complex current $I(t)$ defined by \eqref{VIrep}. %{\color{red} explain the transformation}{\color{green}Done.}.
The values $L = L_c + L_l$  and $R = R_c + R_l$ are the aggregate inductance and resistance of connection as seen by the inverter terminal, respectively. %The instantaneous active and reactive powers can be expressed  as $S=P + j Q= U\mathrm{e}^{j\theta}I^*$ where $I^*$ is the conjugate of $I(t)$. 

With a typical low voltage microgrid in mind, the system parameters shown in Table \ref{T:n5_parameters_App} in the Appendix will be used \cite{nikolakakos2016stability}. For this system, the characteristic electromagnetic time (assuming a 1 km length of connecting line) is $L/R \approx 3.1 ms$, below both the base cycle period of $2\pi\omega_0^{-1} = 20 ms$ and the characteristic timescale of droop controls ($\tau \approx 31.8 ms$). %{\color{red} Is it correct ?}{\color{green}Yes, it is 31 rads/per s, and this is the number which enters the filter}). 
The strong time-scale separation in such a system is usually used as a justification for model order reduction. Indeed, given the fast electromagnetic transients, one may assume that the currents $I_d, I_q$ always remain close to their quasi-stationary values derived from Kirchhoff's laws. Formally, this procedure is equivalent to neglecting the derivative terms in the left hand side of \eqref{ieqD} and \eqref{ieqQ}. %This approximation is universally accepted in the analysis of small-signal stability of synchronous generators in traditional power systems.
This approximation is universally accepted for small-signal stability analysis in traditional power systems. However, in the following discussion, the inappropriateness of using the above approximations is to be demonstrated and investigated. Also, a discussion on the strong effect of electromagnetic transients on microgrid stability will be carried out with the introduction of the proposed reduction procedure for accurate stability assessment.

\subsection{Conventional $3^{rd}$-Order Model}

As discussed above within a traditional quasi-equilibrium approximation, one can neglect the effect of electromagnetic transients which formally corresponds to setting the derivative terms to zero in the left-hand terms of \eqref{ieqD} and \eqref{ieqQ}. The line currents become algebraic, and can be trivially solved for the equilibrium current given by 
\begin{equation}\label{curzero}
 I^0 = (R + j\omega_0 L)^{-1}\left(U \mathrm{e}^{j\theta}- U_s\right)
\end{equation}
where the subscript $\{0\}$ and superscript $\{0\}$ denote the nominal value and the zero-order term, respectively. %For the following derivations we switch to per-unit system with the $U_b$ being the base peak phase-to-ground voltage and $S_b$ being the base apparent power.
For the following derivations, the per-unit system will be utilized with $U_b$ being the base peak phase-to-ground voltage and $S_b$ being the base apparent power.
Then, the following expressions for active and reactive power can be obtained from \eqref{curzero}:
\begin{align} 
P^0 &= B\sin \theta + G (U/U_s-\cos \theta)  \label{Psimple}\\ 
Q^0  &= B(U/U_s - \cos \theta) - G \sin \theta  \label{Qsimple}
\end{align}
where $B = U U_s\omega_0 L/(R^2 + \omega_0^2 L^2)$ and $G =  U U_s R/(R^2 + \omega_0^2 L^2)$
. The small-signal stability of the base operating point will be assessed by introducing the small deviations in the angle $\delta \theta$ and in the %logarithmic 
normalized voltage $\delta \rho = \delta U/U$. Then, the linearized equations can be rewritten in the following form:
\begin{subequations}\label{linearized}
\begin{align}
&    \lambda_p \tau \ddot{ \delta \theta} + \lambda_p \dot{ \delta \theta} + \frac{\partial P^0}{\partial \theta} \delta \theta + \frac{\partial P^0}{\partial \rho} \delta \rho = 0 \\
& \lambda_q \tau \dot {\delta \rho} + \lambda_q  \delta  \rho + \frac{\partial Q^0}{\partial \theta} \delta \theta + \frac{\partial Q^0}{\partial \rho} \delta \rho = 0
\end{align}
\end{subequations}
where $\lambda_p = m_p^{-1}$,  $\lambda_q = n_q^{-1}$, $\tau = w_c^{-1}$, and $\omega_0 = 100\pi$. %It should be addressed that $\delta\rho$, $\delta\theta$, $U$, $U_s$, $U_0$, $S_n$, $G$, and $B$ are all dimensionless in this expression. We then assumed that the operating point is characterized by small angles $\theta \approx 0$ and voltage levels are close to nominal value which we chose to be: $U \approx U_0 \approx U_s= 1~pu$.
It is important to note that $\delta\rho$, $\delta\theta$, $U$, $U_s$, $U_0$, $S_n$, $G$, and $B$ are all dimensionless in this expression. We then assumed that the operating point is characterized by small angles $\theta \approx 0$ and voltage levels are close to nominal value which we chose to be: $U \approx U_0 \approx U_s= 1~pu$.
%For typical parameters that we are using (Table \ref{T:n5_parameters_App}) this assumption is well justified, as the typical angle difference and relative voltage deviations are of the order  $\sim 0.01$. Extensions of the analysis to heavily loaded regimes is straightforward but bulky and will be presented elsewhere. Under these assumptions the system \eqref{linearized} reduces to
For the typical parameters used in this paper, this assumption is well justified, as the typical angle difference and relative voltage deviations are of the order $\sim 10^{-2}$. Extensions of the analysis to heavily loaded regimes is straightforward but bulky and will be presented in subsequent publications. Under these assumptions, the system in \eqref{linearized} reduces to a concise form:
\begin{subequations}\label{linbase}
\begin{align}\label{thetaeqn}
 &    \lambda_p \tau\ddot{ \delta \theta} + \lambda_p \dot{ \delta \theta }+  B \delta \theta +  G \delta \rho = 0 \\
& \lambda_q \tau \dot {\delta \rho }  + (\lambda_q  +  B)\delta \rho -  G \delta\theta = 0 \label{rhoeq}
\end{align}
\end{subequations}
The form of equations in \eqref{linbase} indicate that in the absence of conductance, the dynamics of the angle and voltage deviations become uncoupled and the system is always stable. Active resistance introduces an effective negative feedback to the system and may lead to the loss of stability. The detrimental effect of the conductance on stability can be illustrated using the following informal argument based on the multi-time-scale expansion approach utilized in this work. Equation \eqref{rhoeq} implies that the voltage deviation follows the deviation of the angle with some delay:
\begin{equation} \label{integral}
 \delta \rho(t) =  \frac{ G}{\lambda_q \tau  }  \int_0^\infty \exp\left[- \frac{(\lambda_q +  B) T}{\lambda_q \tau  }\right] \delta\theta(t-T) dT,
\end{equation}
%Whenever the dynamics of $\delta \theta$ can be considered to be slow enough the effect of delay can be effectively approximated as
When the dynamics of $\delta \theta$ are slow enough, the effect of delay can be approximated as
\begin{equation} \label{approx}
 \delta \rho(t) \approx \frac{ G}{\lambda_q +  B}\delta\theta(t) -  \frac{\lambda_q \tau  G}{(\lambda_q +  B)^2} \dot{\delta \theta}(t)
\end{equation}
%This expansion can be obtained by applying the first order Taylor expansion to $\delta\theta(t-T)$ in \eqref{integral} and neglecting the contributions of higher derivatives of  $\delta \theta$. Plugging the expression \eqref{approx} back in \eqref{thetaeqn} we obtain the following approximate reduced dynamics of $\delta \theta$:
This expansion can be obtained by applying a first-order Taylor expansion to $\delta\theta(t-T)$ in \eqref{integral} and neglecting the contribution of higher-order derivatives of $\delta \theta$. Plugging the expression \eqref{approx} back in \eqref{thetaeqn}, the following approximation is obtained:
\begin{equation}\label{resthet}
% \lambda_p \tau\ddot{ \delta \theta} + \left[\lambda_p - \frac{\lambda_q \tau  G^2}{(\lambda_q +  B)^2}\right] \dot{ \delta \theta } + \left( B + \frac{\lambda_q \tau  G^2}{\lambda_q +  B}\right)\delta \theta = 0\nonumber
\lambda_p \tau\ddot{ \delta \theta} + \left[\lambda_p - \frac{\lambda_q \tau  G^2}{(\lambda_q +  B)^2}\right] \dot{ \delta \theta } + \left( B + \frac{G^2}{\lambda_q +  B}\right)\delta \theta = 0
\end{equation}
%This approximate equation illustrates the effect of delay on the system stability. For high enough levels of conductance the effective damping coefficient in front of $\dot{\delta \theta}$ can become negative and lead to loss of stability. This can happen for arbitrary ratio of timescales of the system modes, since the characteristic timescale is not the only relevant parameter but rather it's product with the corresponding gain. Within this approximation, the system would remain stable whenever the droop $m_p$ satisfies
The above approximation illustrates the effect of delay on the system stability. For high conductance values, the effective damping coefficient in front of $\dot{\delta \theta}$ can become negative and result into instability. This can happen for any arbitrary ratio of timescales of the system modes, since the characteristic timescale is not the only relevant parameter but rather it's the product with the corresponding gain. Within this approximation, the system would remain stable whenever   droop $m_p$ satisfies
\begin{equation}\label{naivekq}
 m_p < \frac{(1+ n_q B)^2}{n_q\tau G}
\end{equation}
%This argument is not entirely rigorous since the dynamics of $\delta \theta$ is not necessarily slower than the dynamics of $\delta\rho$, although the resulting condition on $m_p$ is reasonably accurate and highlights the importance of delays. However, the same procedure can be applied to account for delays caused by the line inductance that will be shown to be important under certain conditions. In the case of line inductance delays the application of multi-time-scale expansion procedure is justified since the electro-magnetic delay time is much smaller than the typical time-scale of voltage and angle dynamics. 
This argument is not entirely rigorous since the  $\delta \theta$ dynamics is not necessarily slower than the dynamics of $\delta\rho$, although the resulting condition on $m_p$ is reasonably accurate and highlights the importance of delays. However, the same procedure can be applied to account for delays caused by the line inductance which is shown to be important under certain conditions. In the case of line inductance delays, the application of multi-time-scale expansion is justified since the electro-magnetic delay time is much smaller than the typical time-scale of voltage and angle dynamics.

\subsection{High-Fidelity $3^{rd}$-Order Model}
%When the $R/X$ ratio for a microgrid gets larger the conventional $3^{rd}$ order model becomes inappropriate  because the electromagnetic transients start to play a critical role in the onset of instability despite their short timescale. Mathematically, these transients manifest themselves in the derivative terms in the left hand side of \eqref{ieqD} and \eqref{ieqQ} which can not be fully neglected. Nevertheless, it is possible to account for these transients by deriving an effective $3^{rd}$ order model.
When the $R/X$ ratio of a microgrid gets larger, the conventional $3^{rd}$-order model becomes inappropriate because the electromagnetic transients start to play a critical role in the onset of instability despite their short timescale. Mathematically, these transients manifest themselves in the derivative terms of the left hand side of \eqref{ieqD} and \eqref{ieqQ} which cannot be fully neglected. Nevertheless, it is possible to account for these transients by deriving an effective $3^{rd}$-order model.
%The systematic mathematical procedure that provides a way for accounting for these types of corrections is developed within the so-called multiple-scales analysis with the use of singular perturbation theory \cite{Textbook}. 
%Specifically, systems with small parameter $\varepsilon$ (one of the timescales) which behavior is not regular upon the limit $\varepsilon\to 0$ should be approached using the singular perturbation theory. 
%Here we provide a simple though less rigorous derivation that leads to the same results. 
%In Laplace domain, the equations \eqref{ieqD} and \eqref{ieqQ} can be explicitly solved for $I_d$ and $I_q$ via a first order transfer function
In Laplace domain, \eqref{ieqD} and \eqref{ieqQ} can be explicitly solved for $I_d$ and $I_q$ via a first-order transfer function
\begin{equation}\label{accurate}
 I= \frac{U \mathrm{e}^{j\theta}- U_s}{R + j\omega_0 L +  s L} = \frac{I^0}{1 + s L / (R + j\omega_0 L)}.
\end{equation}

%Whenever the goal is to derive an equivalent reduced order model capturing the dynamics of the slow modes, it is reasonable to assume that $|s L / (R + j\omega_0 L) | \ll 1$, relation that holds for modes that evolve on time-scales slower than the electromagnetic time $L/R$. In this case, one can perform Taylor expansion of the expression \eqref{accurate} and arrive at
Whenever the goal is to derive an equivalent reduced-order model capturing the dynamics of slow modes, it is reasonable to assume that $|s L / (R + j\omega_0 L) | \ll 1$ holds for modes that evolve time-scales slower than the electromagnetic time $L/R$. In this case, one can perform Taylor expansion on \eqref{accurate} to get
\begin{equation} \label{approximate}
 I \approx I^0 - \frac{L s}{R + j\omega_0 L} I^0.
\end{equation}
%Returning back to time-domain one derives in this approximation
Returning back to time-domain, \eqref{approximate} can be rewritten as
\begin{equation}\label{I2}
 I \approx I^0 - \frac{L}{R + j\omega_0 L} \frac{d I^0}{dt}
\end{equation}
%The approximate values of $P$ and $Q$ are derived to be
Then, the approximate values of $P$ and $Q$ are obtained as follows:
\begin{align}
 P &\approx P^0 - G'\dot{\rho} - B'\dot\theta \\
 Q &\approx Q^0 -B' \dot \rho  + G' \dot\theta,
\end{align}
where $G'$ and $B'$ are given by
%\begin{align}\label{GB}
% G' &= \frac{L (R^2- \omega_0^2 L^2)}{(R^2+ \omega_0^2 L^2)^2} \\ \nonumber
% B' &= \frac{2 \omega_0 R L^2}{(R^2+ \omega_0^2 L^2)^2}.
%\end{align}
\begin{equation}\label{GB}
 G' = \frac{L (R^2- \omega_0^2 L^2)}{(R^2+ \omega_0^2 L^2)^2};
\quad  B' = \frac{2 \omega_0 R L^2}{(R^2+ \omega_0^2 L^2)^2} .
\end{equation}
%Dynamic equations instead of \eqref{linbase} become:
Then, the dynamic equations in \eqref{linbase} become:
\begin{subequations}\label{linbase2}
\begin{align}\label{thetaeq2}
 &    \lambda_p \tau\ddot{ \delta \theta} + \left(\lambda_p-B'\right) \dot{ \delta \theta }+  B \delta \theta +  G \delta \rho -G' \dot{\delta \rho} = 0 \\
& \left(\lambda_q \tau   -B'\right) \dot {\delta \rho }  + (\lambda_q   +  B) \delta \rho -  G \delta\theta + G' \dot{\delta\theta} = 0 \label{rhoeq2}
\end{align}
\end{subequations}
%These equations now can be analysed using an integral form of solution similar to \eqref{integral} to obtain a generalization of \eqref{naivekq}. However, some straightforward qualitative conclusions can be made even from the basic structure of equations \eqref{linbase2}. The natural negative feedback terms for $\dot{\delta \theta}$ and $\dot{\delta\rho}$ (second terms in equations \eqref{linbase2}) can change sign when the corresponding droop coefficients are increased - the effect that was not present in quasi-stationary $3^{rd}$ order model. A simple set of stability conditions can thus be obtained by requiring these terms to be positive, i.e. $\left(\lambda_p-B'\right)>0$ and  $\left(\lambda_q \tau -B'\right)>0$ which upon substitution of $\lambda_p$, $\lambda_q$ and $B'$ turns into:
These equations now can be analyzed in a similar way to obtain a generalized version of \eqref{naivekq}. However, some straightforward qualitative conclusions can be made from the basic structure of \eqref{linbase2}. The natural negative feedback terms for $\dot{\delta \theta}$ and $\dot{\delta\rho}$ can change sign when the corresponding droop coefficients are increased - the effect that was not present in the conventional $3^{rd}$-order model. Thus, a simple set of stability conditions can be obtained by requiring these terms to be positive, i.e., $\left(\lambda_p-B'\right)>0$ and $\left(\lambda_q \tau -B'\right)>0$ which upon substitution of $\lambda_p$, $\lambda_q$ and $B'$ %(and switching back from per unit system)
turns into:
\begin{equation}
k_p < S_n \frac{(R^2+X^2)^2}{2RX^2};\quad k_q < \tau \omega_0 \frac{S_n}{U_0} \frac{(R^2+X^2)^2}{2RX^2},\label{kpkq}
%k_p < \frac{S_n}{U_s^2} \frac{(R^2+X^2)^2}{3RX^2};\quad k_q < \tau \omega_0 \frac{S_n}{U_s^2} \frac{(R^2+X^2)^2}{3RX^2},\label{kpkq}
\end{equation}
%We would like to specifically emphasize, that the small timescale of the electromagnetic phenomena $L/R$ can not be used as a reliable indicator of the insignificance of the network dynamics. Specifically, even if the second term in \eqref{I2} is small compared to the first (which is the case and is the base for expansion) it enters the dynamic equation through the term of a different order (derivative terms in \eqref{linbase2}), so that the true conditions on the significance of network dynamics are $B'\ll \lambda_p$ and $B'\ll \tau\lambda_q$ with the former usually being stronger.
where $X=\omega_0 L$ and $k_p$, $k_q$ are the normalized frequency and voltage droop gains, respectively (recall that $k_p = m_p S_n/\omega_0$, $k_q = n_q S_n/U_0$, and $S_n$, $U_0$ are both dimensionless in this expression).  It is important to emphasize that the small timescale of the electromagnetic phenomena $L/R$ cannot be used as a reliable indicator of the insignificance of the network dynamics. Specifically, even if the second term in \eqref{I2} is small compared to the first (which is actually the case and the base for expansion), it enters the dynamic equation through a term of different order (the derivative terms in \eqref{linbase2}), so that the true conditions on the significance of network dynamics are $B'\ll \lambda_p$ and $B'\ll \tau\lambda_q$ with the former being usually stronger. 
%The relations \eqref{kpkq} give a rather general estimation of the stability boundary in terms of frequency and voltage droop coefficients and are very good for demonstrating the trends. The general observations from relations \eqref{kpkq} are:
The relations in \eqref{kpkq} give a rather general estimation of the stability boundary in terms of frequency and voltage droop coefficients and are very good for demonstrating the trends. The general observations from \eqref{kpkq} are:
%\begin{enumerate}
%    \item decreasing the inverter rating i.e. connecting smaller inverter with the same relative settings reduces stability
%    \item decrease in the line reactances and resistances (i.e. strengthening the connection to the grid) has a deteriorating effect on stability
%    \item increasing the inverter control filtering time has no effect on the stability boundary in respect to frequency droop but enhances the stability in respect to voltage droop
%\end{enumerate}
\begin{enumerate}
    \item Decreasing the inverter rating (i.e., connecting smaller inverter with the same relative settings) reduces stability.
    \item Decrease in the line reactances and resistances (i.e., strengthening the connection to the grid) has a deteriorating effect on stability.
    \item Increasing the inverter control filtering time has no effect on the stability boundary with respect to frequency droop but enhances the stability with respect to voltage droop.
\end{enumerate}

%This general stability properties have no analogy on the level of large scale power systems. In fact, the first two of them are exactly the opposite of what is true for transmission grids stability where increasing the grid strength always has a positive effect on stability \cite{machowski2011power}. 
These general stability properties have no analogy on the level of large-scale power systems. In fact, the first two are exactly the opposite of what is truly known for transmission grids where increasing the grid strength always has a positive effect on stability \cite{machowski2011power}.

\begin{figure}[t!]
\centering
        \includegraphics[width=0.35\textwidth]{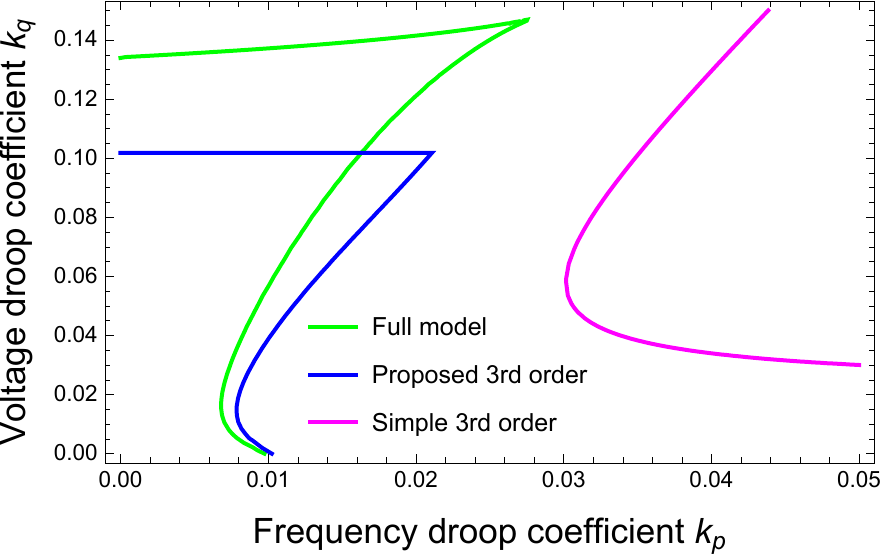}
        %\caption{Stability regions found using three different models}
        %\caption{A comparison of stability regions for three different models - full 5th order model, simple 3rd order model and high fidelity 3rd order model.}
        \caption{A comparison of stability regions for three different models.}
        \label{fig:orders}
\end{figure}

%The comparison of the three models: formal exact 5th order, conventional 3rd order and high-fidelity 3rd order is presented on the Fig. \ref{fig:orders} with the stable region being to the left of each curve. It is obvious that the conventional 3rd order model is highly inappropriate for stability assessment predicting a stability region considerably larger than the exact and high fidelity $3^{rd}$ order model.
A comparison of three different models (the full $5^{th}$-order model presented in \eqref{thetaeq}--\eqref{ieqQ}, the conventional $3^{rd}$-order model and the proposed high-fidelity $3^{rd}$-order model) is presented in Fig. \ref{fig:orders} with the stable region being to the left of each curve. It is obvious that the conventional $3^{rd}$-order model is highly inappropriate for stability assessment where it predicts a larger stability region than the other two models.

The numerical simulation using a \SI{10}{\kilo VA} inverter connected to a grid through a line (parameters are taken from Table \ref{T:n5_parameters_App} in the Appendix) gives a stability boundary of $k_p\sim0.5-2\%$ and $k_q\sim2-25\%$ depending on the connecting line length and filter time constant. The result is specific to microgrids and has no analogy to large-scale transmission grids, and can be understood in the following way. Let us use a ``line rating'' quantity $S_l\sim V^2/Z$ which characterise the formal angular or voltage stability boundary for the line. Then, according to \eqref{kpkq}, the maximum value of droop coefficient is simply the ratio of inverter rating to line rating. For the parameters under consideration, the line rating is of the order of several hundreds of kilowatts which is two orders of magnitude higher than the typical inverter rating. Contrary to large transmission systems, where power flows are mostly limited by voltage drop and angular stability, the main limitation in microgrids is the overcurrent limit in conductors. Consequently, microgrids typically operate in a region of very small values of inverter angles $\theta$ (or, more precisely, angle differences), this fact was also noted in \cite{guo2014dynamic}. For large transmission systems, generator ratings are usually of the same order as line ratings (mainly due to machine internal inductances) and, hence, the formal stability limit for machine is around $k_p \sim 100\%$ which is never used in practice for other reasons. It is therefore, the absence of large internal impedance which makes the inverters completely different to synchronous machines in terms of stability. A synchronous machine connected to a low-voltage grid doesn't also exhibit this type of instabilities since machines always have large internal reactance $X'\sim 0.2-0.5$ which effectively weaken the grid. From this point of view, one can also give a rather simple explanation why the electromagnetic transients are not important for large-scale power systems (despite them having larger timescale due to more inductive lines). The effect is negligible if the $B'$ term is much smaller than $\lambda_p$. The former has an order of magnitude similar to the inverse impedance in p.u. which for large-scale power grids is around unity, while the latter is the inverse frequency droop - at least one order of magnitude higher. It is also noted that these effects have no relation to generator time constant or, in case of inverter, filter time constant $\tau$, such that it is the small per-unit value of network characteristic impedance (``strength'' of the grid) which makes it necessary to consider electromagnetic transients.    

%Note: \label{} should be after \caption{}%

\begin{figure}[t!]
\centering
        \includegraphics[width=0.35\textwidth]{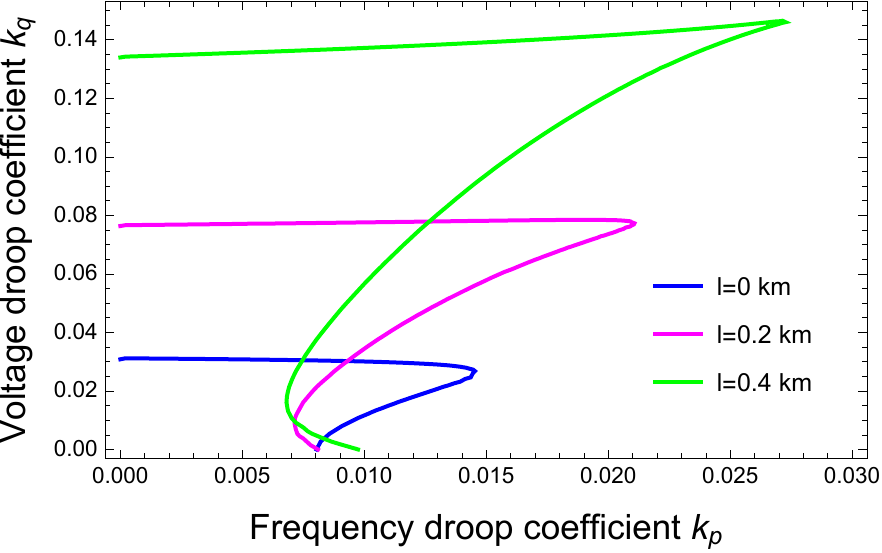}
        %\caption{Stability regions depending on the connection line length - shortening the connection lines makes stability region smaller.  For all cases there is additional coupling impedance between inverter terminal and grid. }
        \caption{Stability regions for different lengths of connection line.}
        \label{fig:stablines}
\end{figure}

\begin{figure}[t!]
\centering
        \includegraphics[width=0.35\textwidth]{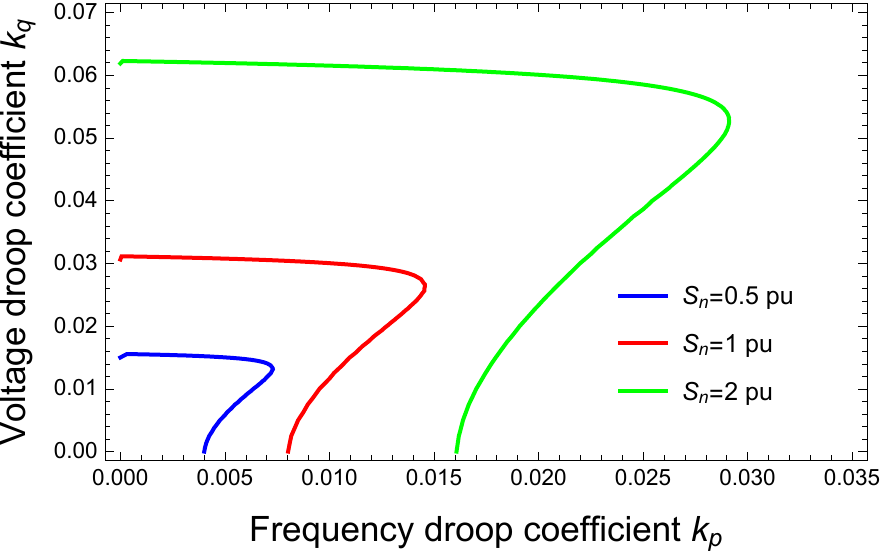}
        %\caption{Stability regions depending on the inverter rating, 1 pu=10 kVA. Smaller inverters have smaller stability regions.} 
        \caption{Stability regions for different inverter rating values, 1 pu =10 kVA.} 
        \label{fig:ratings}
\end{figure}

%The influence of the connecting line length on stability is illustrated on the Fig. \ref{fig:stablines} with the blue curve corresponding to direct inverter connection where the effective line impedance is only due to an internal coupling. As was noticed above, the increase in the connecting line impedance tends to increase the stability region along the vertical axis. Since there is always a coupling impedance present one can think about the existence of a robust stability region for a given inverter corresponding to the lower left corner of Fig. \ref{fig:stablines}. \textcolor{red}{(Mike: After discussion, Petr also thinks we need to put more explanation to avoid confusions)} Nevertheless, the situation is different if one considers the influence of the inverter rating which is illustrated by Fig. \ref{fig:ratings} for inverters of $10$, $7.5$ and $5$ kVA respectively. The stability criteria for small inverter are becoming rather strict with the acceptable values of frequency droop $k_p$ becoming less than $1\%$. One important practical conclusion from this fact is that connection of few smaller inverters instead of a large one can deteriorate the stability of the system. 
The influence of different connecting line lengths on stability is illustrated in Fig. \ref{fig:stablines} with the blue curve corresponding to direct inverter connection and the effective line impedance is only due to the internal coupling impedance. As noted in Fig. \ref{fig:stablines}, the increase in the connecting line impedance tends to increase the overall stability region especially in terms of voltage droop coefficient. While there is no strict monotonic dependence of the maximum frequency droop coefficient on the connecting line lengths, there seems to exist a robust stability region corresponding to the lower left corner of Fig. \ref{fig:stablines} which is due to the minimum coupling impedance always being present in the system. Nevertheless, the situation is different if the influence of inverter rating is considered, and this is illustrated in Fig. \ref{fig:ratings} for inverters of $5$, $10$ and \SI{20}{ \kilo VA} ratings, respectively. The stability criteria for small inverters are becoming rather strict with the acceptable values of frequency droop $k_p$ becoming less than $0.5\%$. One important practical conclusion from this fact is that the connection of few smaller inverters instead of a large one can deteriorate the stability of the system. 

Let us now give a rather simple physical interpretation to the instability mechanism in terms of time delays in current. One can think about the exact current $i(t)$ being retarded with respect to quasi-stationary value $i^0$ by the characteristic electromagnetic time $L/R$ which decreases as $R$ increases, such that one might expect the quasi-stationary approximation (conventional $3^{rd}$-order model) to work better. However, it is not the delay itself, but rather the product of delay and gain that determines the overall effect on stability. While the delay time is inversely proportional to $R$, the gain, which is determined by $1/B'$, is proportional to $R^2$ and, hence, the quasi-stationary approximation becomes invalid despite the decrease in electromagnetic delay. 

%----------------------------------------------------%
%               S E C T I O N  III
%----------------------------------------------------%      

\section{Generalized Multi-Timescale Approach}\label{sec:genap}

%In this section we present a general formulation of the method for analysis of stability for systems with multiple time-scales. The method represents the first order of the singular perturbation theory as opposed to zeroth order, which corresponds to neglecting the dynamics of fast variables altogether. Employing this method allows for proper inclusion of possible fast variables dynamics on slow modes. The presence of strong timescale separation in microgrids manifests itself in the appearance of several clusters of modes on the plane of system eigen-values with only one cluster - the slowest one, associated with power controls, being of interest from the point of view of small-signal stability. \cite{mohamed2008adaptive,pogaku2007modeling}. We start from the general description of a system with a set of first order differential equations linearized around a supposed equilibrium point:
In this section, a general formulation method for stability analysis of multiple time-scales systems is presented. The method represents a first-order of the singular perturbation theory as opposed to zero-order, which corresponds to neglecting the dynamics of fast variables altogether. Employing this method allows for proper inclusion of possible fast variables dynamics on slow modes. The presence of strong timescale separation in microgrids manifests itself in the appearance of several clusters of modes on the plane of system eigen-values with only one cluster - the slowest one, associated with power controllers, which are of main interest from the point of view of small-signal stability \cite{mohamed2008adaptive,pogaku2007modeling}. Let us start from the general description of a system with a set of first-order differential equations linearized around an equilibrium point:
\begin{equation}
\delta{\dot x}=A\delta x    \label{initial}
\end{equation}
%where $x$ is a set of system variables and $A$ is the corresponding Jacobian matrix. It is desirable to aim at such a simplification of a system representation, so that only the relevant modes are considered in the form of dynamic equations and all the rest are properly eliminated. The timescale separation was appreciated in \cite{nikolakakos2016stability} where the authors introduced two time-scale model of a system and completely excluded the dynamics of "fast" variables by using their quasi-stationary values and considered three different ways of separation between "fast" and "slow" degrees of freedom. In the present paper we perform a more systematic procedure of timescale separation and describe a procedure for exclusion of fast degrees of freedom with their effect properly reflected in the reduced order system. We start from separating the system \eqref{initial} into two subsystems corresponding to slow and fast variables respectively (we use denotations from \cite{nikolakakos2016stability}):
where $x$ is a set of system variables and $A$ is the corresponding Jacobian matrix. It is desirable to aim at such a simplification of a system representation, so that only the relevant modes are considered in the form of dynamic equations and all the rest are properly eliminated. The timescale separation was presented in \cite{nikolakakos2016stability} where the authors introduced a two time-scale model of a system and completely excluded the dynamics of ``fast'' variables by using their quasi-stationary values and considered three different ways of separation between ``fast'' and ``slow'' degrees of freedom. In this paper, a more systematic procedure of timescale separation will be presented along with a procedure for proper exclusion of fast degrees of freedom while accounting for their effect in the reduced-order system.The separation of the system in \eqref{initial} into two subsystems corresponding to slow and fast variables gives: %(denotations from \cite{nikolakakos2016stability} are used)
\begin{equation}
\delta{\dot x_{s}}=A_{ss}\delta x_s + A_{sf}\delta x_f    \label{slow}
\end{equation}
\begin{equation}
\Gamma \delta{\dot x_{f}}=A_{fs}\delta x_s + A_{ff}\delta x_f   \label{fast}
\end{equation}
%where the subscripts $s$ and $f$ correspond to slow and fast degrees of freedom, $\Gamma$ is a set of parameters designating fast degrees of freedom. It is reasonable to consider a general case where $\Gamma$ is a matrix, i.e. dynamic equations for the fast degrees of freedom are written in a general form without resolving them in respect to individual first derivatives and can also include the algebraic equations for which the corresponding components of $\Gamma$ are zero. A procedure employed in \cite{nikolakakos2016stability} reduces to simply neglecting the left-hand side of \eqref{fast}, thus reducing the systems for $x_s$ to the following:
where the subscripts $s$ and $f$ correspond to slow and fast degrees of freedom, respectively; $\Gamma$ is a set of parameters designating fast degrees of freedom. %\textcolor{blue}{(deleted)}It is reasonable to consider a general case where $\Gamma$ is a matrix, i.e., the dynamic equations for fast degrees of freedom are written in a general form without resolving them with respect to individual first derivatives and can also include the algebraic equations for which the corresponding components of $\Gamma$ are zero.
A procedure employed in \cite{nikolakakos2016stability} neglects the left-hand side of \eqref{fast}, thus reducing the system in \eqref{slow} to the following:
\begin{equation}\label{firstorder1}
\delta{\dot x_{s}}=(A_{ss}- A_{sf}A^{-1}_{ff}A_{fs})\delta x_s 
\end{equation}

%The expression \eqref{firstorder1} can be treated as a zeroth order approximation in the perturbation expansion. It is formally obtained by stating a linear relation between $\delta x_f$ and $\delta x_s$ which is found from equation \eqref{fast} by neglecting it's right-hand side. Let us now consider the next order by stating that $\delta x_f$ should also be linearly dependant on the first derivatives of $\delta x_s$ i.e. on $\dot{\delta x_s}$. Inserting such a dependence in \eqref{fast} and separating different orders of magnitude one finds:
Expression \eqref{firstorder1} can be treated as a zero-order approximation in the perturbation expansion. It is formally obtained by stating a linear relation between $\delta x_f$ and $\delta x_s$ which is found from \eqref{fast} by neglecting its left-hand side. Let us now consider the next order by stating that $\delta x_f$ should also be linearly dependant on the first derivative of $\delta x_s$ (i.e., on $\dot{\delta x_s}$). Inserting such a dependence in \eqref{fast} and separating different orders of magnitude, one finds:
\begin{equation}
\delta x_{f}=-A^{-1}_{ff} A_{fs} \delta x_s - A^{-1}_{ff} \Gamma A^{-1}_{ff} A_{fs} \delta \dot{x_s}
\end{equation}
%Inserting this into equation \eqref{slow} one finds:
Inserting this into \eqref{slow}, the following is obtained:
\begin{equation}\label{robust}
(1+A_{sf}A^{-1}_{ff} \Gamma A^{-1}_{ff}A_{fs})\delta{\dot x_{s}}=(A_{ss}- A_{sf}A^{-1}_{ff}A_{fs})\delta x_s 
\end{equation}
%which is a generalization of equations \eqref{firstorder1} and $1$ in the left-hand side is a unity matrix. The described procedure is rather general and incorporates the cases when some of the fast degrees of freedom are "instantaneous" which correspond to respective elements of $\Gamma$ matrix being zero so that algebraic constraints are also incorporated. The convenience of the representation used lies in the fact that one can operate with a general set of fast degrees of freedom without the need to first separate the linear independent ones or resolve for individual variables derivatives.
which is a generalization of \eqref{firstorder1} and $1$ in the left-hand side of \eqref{robust} is a unity matrix. The described procedure is rather general and incorporates the cases when some of the fast degrees of freedom are ``instantaneous'' which correspond to respective elements of $\Gamma$ being zero such that algebraic constraints are also considered. The convenience of the representation used lies in the fact that one can operate with a general set of fast degrees of freedom without the need to first separate the linear independent ones or solve for individual variables derivatives.

%----------------------------------------------------%
%               S E C T I O N  IV
%----------------------------------------------------%    

\section{Network Generalization and Stability Certificates}\label{net}

%Generalization of the method to networks is done directly by constructing a system of dynamic equations similar to \eqref{linbase2} for every inverter node. First, a network admittance matrix $\mathbf{Y}$ should be constructed using the full network impedance matrix where all the line and load impedances  $Z_{ij}$ are written in Laplace domain i.e. $Z_{ij}=R_{ij}+j\omega_0 L_{ij}+sL_{ij}$. Matrix $\mathbf{Y}$ relates inverter voltages with inverter currents:
Generalization of the proposed model presented in Section \ref{sec:2bus} to networks is done directly by constructing a system of dynamic equations similar to \eqref{linbase2} for every inverter node. First, a network admittance matrix $\mathbf{Y}$ should be constructed using the full network impedance matrix where all the line and load impedances  $Z_{ij}$ are written in Laplace domain (i.e., $Z_{ij}=R_{ij}+j\omega_0 L_{ij}+sL_{ij}$). Matrix $\mathbf{Y}$ links inverter voltages to inverter currents:
\begin{equation}\label{Ys}
\mathbf{I}(s)=\mathbf{Y}(s)\mathbf{V}(s)   \end{equation}
%where $\mathbf{I}(s)$ and $\mathbf{V}(s)$ are Laplace transforms of (complex) vectors of inverter currents and voltages. The equivalent network contains inverter buses connected with each other through lines in a certain way as well as effective shunt elements attached to inverter buses representing loads. It is convenient to separate the total admittance matrix into  ``network'' (denoted by index $N$) and ``load'' (denoted by index $L$) parts:
where $\mathbf{I}(s)$ and $\mathbf{V}(s)$ are the Laplace transforms of the complex vectors of inverter currents and voltages, respectively. The equivalent network contains inverter buses that are interconnected through connection lines in addition to shunt elements attached to inverter buses to represent loads. It is convenient to separate the total admittance matrix into ``network'' (denoted by index $N$) and ``load'' (denoted by index $L$) parts:
\begin{equation}\label{YNL}
\mathbf{Y}(s)=\mathbf{Y}_N(s) + \mathbf{Y}_L(s)   \end{equation}
%We note, that the ``load'' admittance matrix $\mathbf{Y}_L$ is diagonal. The next step is to expand the admittance matrix in series of Laplace parameter $s$ up to the first order:
where the ``load'' admittance matrix $\mathbf{Y}_L$ is diagonal. Then, the next step is to expand the admittance matrix using first-order Taylor expansion:
\begin{equation}\label{Ymat}
\mathbf{Y}(s)\approx \mathbf{Y}_0 + \mathbf{Y}_1s    
\end{equation}
where 
\begin{equation}
\mathbf{Y}_0=\mathbf{Y}(s)|_{s=0}
\end{equation}
\begin{equation}
\mathbf{Y}_1=\frac{\partial \mathbf{Y}(s)}{\partial s}|_{s=0}
\end{equation}
%Inserting it to \eqref{Ys} and switching back to time domain we get:
After substitution in \eqref{Ys} and switching back to time domain, a generalized version of \eqref{I2} is obtained:
\begin{equation}\label{Inet}
\mathbf{I}(t)=\left[\mathbf{Y}_{0N}+ \mathbf{Y}_{0L}\right]\mathbf{V}(t) + \left[\mathbf{Y}_{1N}+\mathbf{Y}_{1L}\right] \dot{\mathbf{V}}(t)    
\end{equation}
%a generalization of \eqref{I2}. We specifically note, that it is not appropriate to use the quasi-stationary Kron reduced admittance matrix (being $\mathbf{Y_0}$) for network dynamics simulation, since the proper network representation for this purpose should be calculated using the initial structure with full impedances (including Laplace parameter $s$).  
One can note that it is not appropriate to use the quasi-stationary reduced admittance matrix ($\mathbf{Y_0}$) for network dynamic simulation, since the proper network representation should be calculated using the initial structure with full impedances (including the Laplace parameter $s$).  

%The relations \eqref{Inet} and \eqref{Ymat} now can be used in order to construct the system of dynamic equations generalizing \eqref{linbase2} for a system of interconnected inverters with loads:
Then, the relations \eqref{Ymat} and \eqref{Inet} can be used to construct the generalized dynamic equations of a system with interconnected inverters and loads and, hence, \eqref{linbase2} become:
%\begin{subequations}\label{veclinear}
%\begin{align} \label{thetaeqvec}
%    &    \tau\mathbf{\Lambda_p} \ddot{\vartheta} + (\mathbf{\Lambda_p} - \mathbf{B'}) \dot\vartheta + \mathbf{B} \vartheta + \mathbf{G} \varrho + \tilde{\mathbf{G}} \varrho - \mathbf{G'} \dot\varrho= 0  \\
%&   (\tau \mathbf{\Lambda_q} -\mathbf{B'})\dot \varrho + \left(\mathbf{\Lambda_q} + \mathbf{B} \right + \tilde{\mathbf{B}}) \varrho - \mathbf{G} \vartheta  + \mathbf{G'} \dot \vartheta = 0 \label{rhoeqvec}
%\end{align}
%\end{subequations}
\begin{subequations}\label{veclinear}
\begin{align} \label{thetaeqvec}
&\tau\mathbf{\Lambda_p} \ddot{\vartheta} + (\mathbf{\Lambda_p} - \mathbf{B'}) \dot\vartheta + \mathbf{B} \vartheta + (\mathbf{G} + \tilde{\mathbf{G}}) \varrho - \mathbf{G'} \dot\varrho= 0  \\
&(\tau \mathbf{\Lambda_q} -\mathbf{B'})\dot \varrho + (\mathbf{\Lambda_q} + \mathbf{B} + \tilde{\mathbf{B}}) \varrho - \mathbf{G} \vartheta  + \mathbf{G'} \dot \vartheta = 0 \label{rhoeqvec}
\end{align}
\end{subequations}
%where now $\vartheta$ and $\varrho$ are vectors of inverter angles and (relative) voltages respectively and all the terms in bold are square matrices with the number of dimensions corresponding to the number of inverters in the grid. The terms $\mathbf{\Lambda_p}$ and $\mathbf{\Lambda_q}$ represent the diagonal matrices with the elements equal to  inverse frequency and voltage droop coefficients respectively.  
where $\vartheta$ and $\varrho$ are vectors of inverter angles and (relative) voltages, respectively; and all the terms in bold are square matrices with dimensions corresponding to the number of inverters in the grid. $\mathbf{\Lambda_p}$ and $\mathbf{\Lambda_q}$ represent the diagonal matrices with elements equal to the inverse of frequency and voltage droop coefficients, respectively.
%Matrices $\mathbf{B}$, $\tilde{\mathbf{B}}$, $\mathbf{G}$ and $\tilde{\mathbf{G}}$  can be expressed in terms of the quasi-stationary network admittance matrix: 
Matrices $\mathbf{B}$, $\tilde{\mathbf{B}}$, $\mathbf{G}$ and $\tilde{\mathbf{G}}$  can be expressed in terms of the quasi-stationary network admittance matrix: 
%\begin{equation}\label{BGmat}
%\mathbf{B}=-U_s^2\textrm{Im}\left\{\mathbf{Y}_{0N}\right\},\qquad \mathbf{G}=U_s^2\textrm{Re}\left\{\mathbf{Y}_{0N}\right\}    
%\end{equation}
%\begin{equation}\label{BGshuntmat}
%\tilde{\mathbf{B}}=-U_s^2\textrm{Im}\left\{\mathbf{Y}_{0L}\right\},\qquad \tilde{\mathbf{G}}=U_s^2\textrm{Re}\left\{\mathbf{Y}_{0L}\right\}    
%\end{equation}
\begin{equation}\label{BGmat}
\mathbf{B}=-U_0^2\textrm{Im}\left\{\mathbf{Y}_{0N}\right\},\qquad \mathbf{G}=U_0^2\textrm{Re}\left\{\mathbf{Y}_{0N}\right\}    
\end{equation}
\begin{equation}\label{BGshuntmat}
\tilde{\mathbf{B}}=-2U_0^2\textrm{Im}\left\{\mathbf{Y}_{0L}\right\},\qquad \tilde{\mathbf{G}}=2U_0^2\textrm{Re}\left\{\mathbf{Y}_{0L}\right\}    
\end{equation}
%We note that both $\mathbf{B}$ and  $\mathbf{G}$ are singular, but positive semi-definite, and matrices $\tilde{\mathbf{B}}$ and $\tilde{\mathbf{G}}$ are diagonal and positive-definite.  
%Matrices $\mathbf{B}'$ and $\mathbf{G}'$ represent the effect of the network and load dynamics (there is no need to make a separation between them for these terms) and can be expressed in terms of $\mathbf{Y}_1$:
It is important to note that both $\mathbf{B}$ and $\mathbf{G}$ are singular but positive semi-definite matrices, while $\tilde{\mathbf{B}}$ and $\tilde{\mathbf{G}}$ are diagonal and positive-definite matrices. Matrices $\mathbf{B}'$ and $\mathbf{G}'$ represent the effect of network and load dynamics, and can be expressed in terms of $\mathbf{Y}_1$:
%\begin{subequations}
%\begin{align}
%&\mathbf{B}'=U_s^2\textrm{Im}\left\{\mathbf{Y}_{1N} + \mathbf{Y}_{1L}\right\}\\
%&\mathbf{G}'=-U_s^2\textrm{Re}\left\{\mathbf{Y}_{1N} + \mathbf{Y}_{1L}\right\}
%\end{align}
%\end{subequations}
\begin{subequations}
\begin{align}
&\mathbf{B}'=U_0^2\textrm{Im}\left\{\mathbf{Y}_{1N} + \mathbf{Y}_{1L}\right\}\\
&\mathbf{G}'=-U_0^2\textrm{Re}\left\{\mathbf{Y}_{1N} + \mathbf{Y}_{1L}\right\}
\end{align}
\end{subequations}
%Since matrices $\mathbf{B}'$ and $\mathbf{G}'$ are obtained from the admittance matrix through linear operation, they preserve the general property: diagonal element is equal to the negative sum of all the elements in a corresponding row plus the shunt admittance due to a load attached to the corresponding bus. We also note, that matrix $\mathbf{B}'$ is positive definite, while matrix $\mathbf{G}'$ is sign indefinite. 
Since $\mathbf{B}'$ and $\mathbf{G}'$ are obtained from the admittance matrix through linear operation, they preserve the general property: diagonal element is equal to the negative sum of all elements in a corresponding row plus the shunt admittance due to a load attached to the corresponding bus. One can also note that matrix $\mathbf{B}'$ is positive definite, while matrix $\mathbf{G}'$ is sign indefinite.
%Equations \eqref{veclinear} allow one to analyse the stability of a multi-inverter grid taking into account the network dynamics, while still having the effective low order form and droop coefficients entering the equation in a simple way. This makes it possible to derive a certain number of local (i.e. containing characteristics of one inverter in each relation) criteria through construction of a Lyapunov function and applying condition of it's decay. This can be done at the expense of being conservative and special approach is needed to choose the proper type of Lyapunov function which will be presented in subsequent publications. 
Equations \eqref{veclinear} allow one to analyze the stability of a multi-inverter system taking into account the network dynamics, while still having an effective low-order form with simple representation of droop coefficients. Therefore, it is possible to derive a certain number of local (i.e., containing characteristics of one inverter in each relation) criteria through the construction of a Lyapunov function and applying conditions of its decay. This can be done at the expense of being conservative and special approach is needed to choose the proper type of Lyapunov function which will be presented in future work. 

%----------------------------------------------------%
%               S E C T I O N  V
%----------------------------------------------------% 

\section{Numerical Evaluation}\label{sec:num}
\subsection{Model Accuracy}

\begin{figure}[t!]
\centering
        \includegraphics[width=0.32\textwidth]{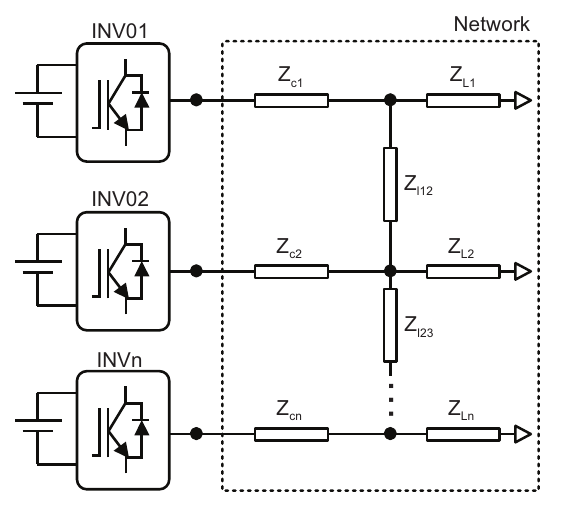}
        %\caption{Comparison of the dynamic responses between different models ( $k_p=0.75\%$).}
        \caption{System configuration of inverter-based microgrid under study.}
        \label{network}
\end{figure}

\begin{figure}[t!]
\centering
        \includegraphics[width=0.4\textwidth]{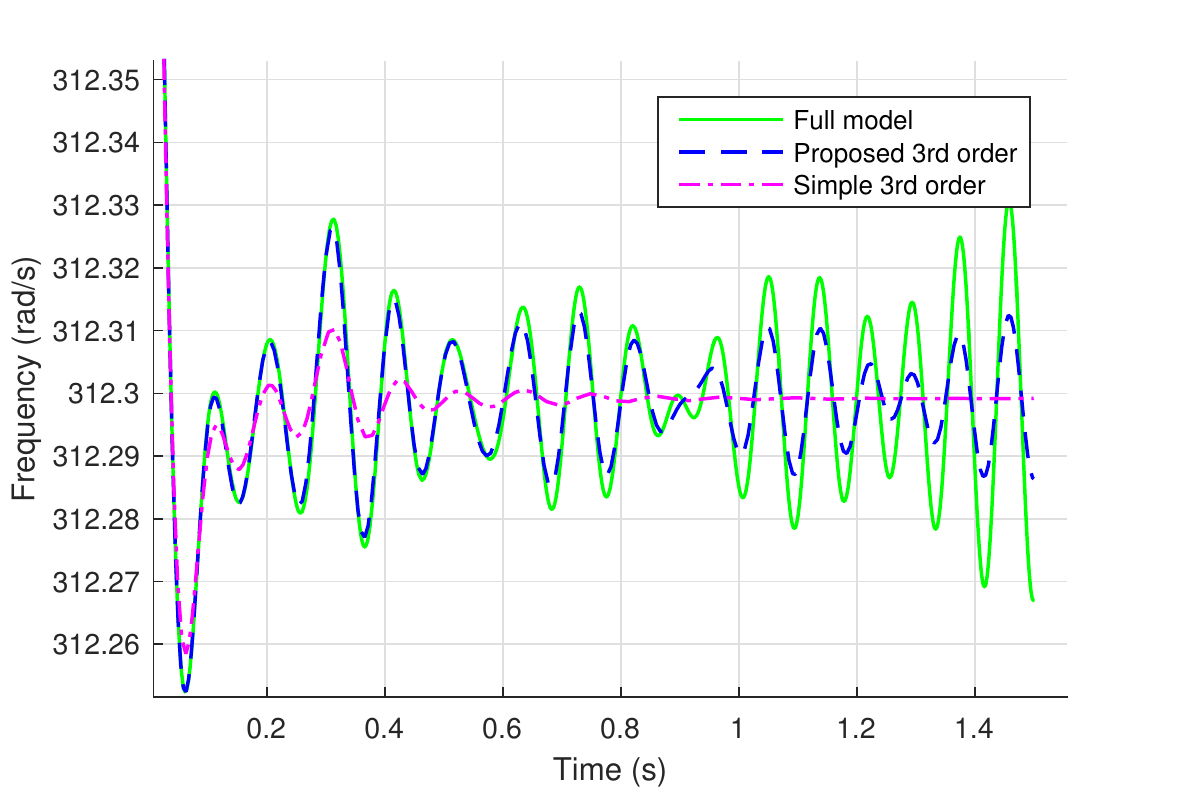}
        \caption{Dynamic responses of different models ($k_p=0.75\%$).}
        \label{time_domain}
\end{figure}

\begin{figure}[t!]
\centering
        \includegraphics[width=0.4\textwidth]{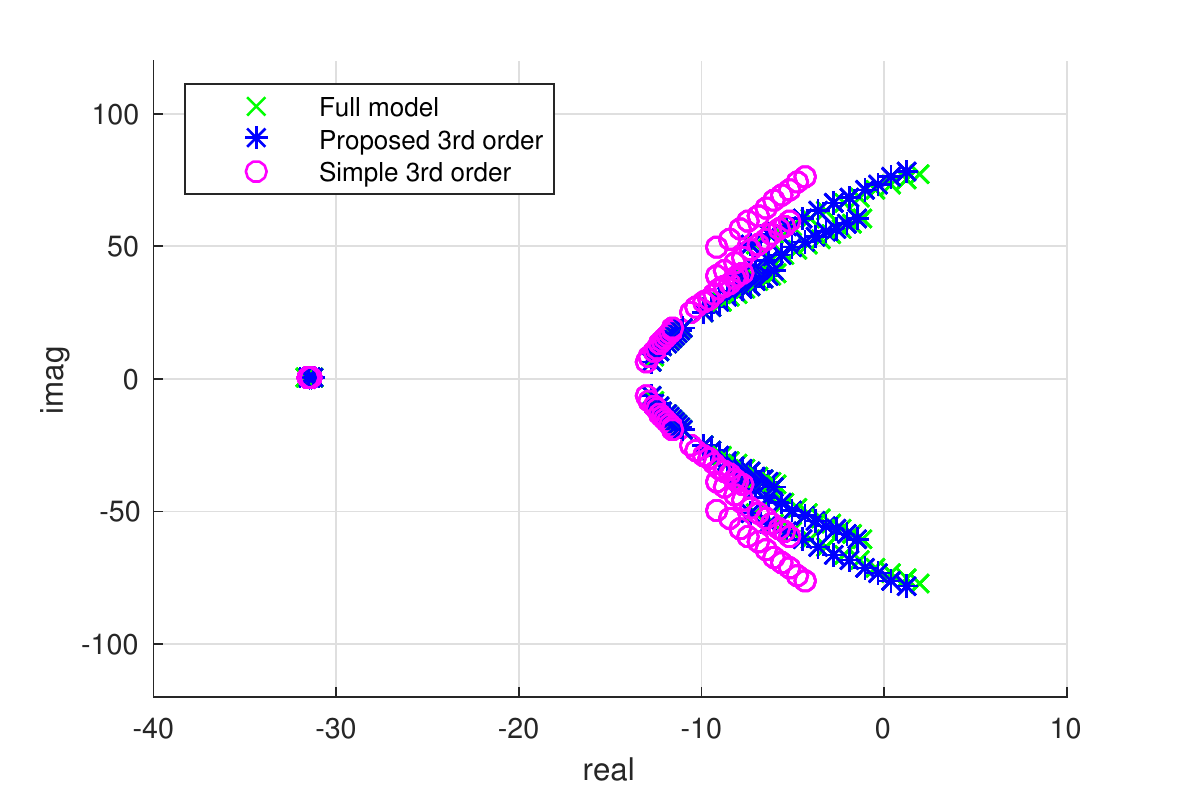}
        %\caption{Comparison of the dynamic responses between the electromagnetic and derived 3rd order models ($k_p=0.3\%$ to $0.75\%$).}
        \caption{Eigenvalue plots of different models ($k_p = 0.3\%- 0.75\%$).}
        \label{pzmap}
\end{figure}

In this section, simulation results comparing the different models are presented. To verify the accuracy of the proposed reduced-order model, a system with five inverters in the cascade configuration shown in Fig. \ref{network} is investigated, in which the coupling inductors are included into the network and $\mathbf{Y}$ representation. The system parameters of five inverter-based microgrid are given in Table \ref{T:n5_parameters_App} in the Appendix. First, a time-domain simulation was conducted to compare the dynamic responses between the different models, as shown in Fig. \ref{time_domain}. The active power droop gain, $k_p$, is chosen to destabilize the system such that erroneous prediction can be observed from the conventional or simple $3^{rd}$-order model. Furthermore. a comparison of eigenvalue movements by varying $k_p$ for different models is given in Fig. \ref{pzmap}. It can be seen that the eigenvalues of the system calculated using the proposed $3^{rd}$-order model are far closer to the full model compared to the simple $3^{rd}$-order model, which is consistent with the simplified two-bus results presented in Section \ref{sec:2bus}. To gain a more effective way of analyzing the prediction errors, the critical $k_p$ is utilized to evaluate the model accuracy. Since the accuracy of the proposed model relies on the fast relaxation of electromagnetic dynamics, a lower $X/R$ ratio, in general, leads to higher accuracy. This is very important as the instability is mainly caused by the low $X/R$ ratio that disrupts the $P-\omega$ and $Q-V$ relations. That is, the proposed model can actually achieve better performance for scenarios with higher resistive network. To verify that, the average prediction errors of both $3^{rd}$-order models as compared to the full model are summarized in Table \ref{T:criticalkp} by varying the average $X/R$ ratio. From Table \ref{T:criticalkp}, it can be seen that the prediction accuracy of the proposed $3^{rd}$-order model is much better than that of the simple one. This further justifies the effectiveness of the proposed reduction technique.

\subsection{Simulation Efficiency}
%Another important feature of the proposed reduced-order is that it reduces the computation burdens of the time-domain simulation. For the full model, all the cable and load dynamics are to modeled as states. The total number of states is approximately 9 times the number of inverters in the cascade topology. However, the proposed technique requires only 3 states per inverter, which means it reduces the number of the states by two-third. This allows us to handle the network systems with a large number of inverters. To identity the efficiency of the proposed model, the full and proposed 3rd order models are tested via the time-domain simulation with the Matlab default O.D.E. solvers. The inverters, coupling inductoes, and the lines/cables are assumed to be identical for simplicity. The simulation time is set to be a second. The results are shown in Table \ref{T:Computation}. The timing procedure was done using a personal computer equipped with Intel Core i7 4790k and 16GB memory. From the results, it is clearly demonstrated that the proposed model reduces the number of states and improves the simulation efficiency significantly. 
Another important feature of the proposed reduced-order is that it reduces the computation burden on the time-domain simulation. For the full model, all the cable and load dynamics are modelled as states. The total number of states ($n_s$) is approximately 9 times the number of inverters in the cascade topology. In comparison, the proposed technique requires only 3 states per inverter, which reduces the number of states by two-third. This allows us to handle a network system with large number of inverters. To identity the efficiency of the proposed model, the full and proposed $3^{rd}$ order models are tested via time-domain simulation with Matlab default O.D.E. solvers. The inverters, coupling inductors, and the lines/cables are assumed to be identical for simplicity. The simulation time is set to be one second. The results are shown in Table \ref{T:Computation} for 5 and 25 inverter-based microgrids. %The timing procedure was done using a personal computer equipped with Intel Core i7 4790k and 16GB memory. 
From the results, it is clearly demonstrated that the proposed model reduces the number of states and improves the simulation efficiency significantly. 

\section{Conclusion}
%Network dynamics in microgrids can greatly influence the behavior of the slow degrees of freedom associated with inverter power controllers. Particularly, the stability region in terms of voltage and frequency droop coefficients is significantly diminished compared to a simple quasi-stationery model. In the present paper we provided an insight to the physical mechanism of instability and presented a method for proper exclusion of fast network degrees of freedom without compromising the accuracy of the model while bringing major simplifications in terms of computational complexity. We used the model to illustrate the microgrid specific effects, namely deterioration of stability by reduction of network impedances and/or inverter ratings. 

%The further studies will focus on the development of stability assessment methods based on the reduced order model presented. The method of Lyapunov functions may allow for formulation of stability criteria dealing with each inverter droop coefficients and connecting lines separately or with pairs of interconnected inverters. Such criteria can be used for assessment of stability during system reconfiguration or multiple microgrids interconnection. 

Network dynamics in microgrids can greatly influence the behavior of slow degrees of freedom associated with inverter power controllers. Particularly, the stability region in terms of voltage and frequency droop coefficients is significantly diminished compared to a simple quasi-stationery model. In this paper, an insight to the physical mechanism of instability is presented along with a method for proper exclusion of fast network degrees of freedom without compromising the accuracy of the model while bringing major simplifications in terms of computational complexity. The influence is reflected in the corresponding change of the coefficients of the resulting third order model compared to a purely quasi-stationery approximation (neglecting the fast degrees of freedom altogether) which leads to significant changes in the predicted regions of stability. The proposed technique is used to illustrate the microgrid specific effects, namely deterioration of stability by reduction of network impedances and/or inverter ratings. Future studies will focus on the development of stability assessment methods based on the proposed reduced-order model. The method of Lyapunov functions may allow for formulation of stability criteria dealing with each inverter droop coefficients and connecting lines separately or with pairs of interconnected inverters. Such criteria can be used for assessment of stability during system reconfiguration or multiple microgrids interconnection.

%\section{Appendix}
%\appendix
%The system parameters are obtained from \cite{pogaku2007modeling} and \cite{nikolakakos2016stability}, and are given in Table \ref{T:n5_parameters_App} for a five inverter-based microgrid.

\begin{table}[t!]
\caption{Parameters of Five Inverter-Based Microgrid }
\label{T:n5_parameters_App}
\begin{center}
\begin{tabular}{c | c | c}
\textbf{Parameter} & \textbf{Description} & \textbf{Value}\\ \hline
$U_{b}$ & Base Peak Phase Voltage & \SI{381.58}{\volt}\\
$S_b$ & Base Inverter Apparent Power & \SI{10}{\kilo VA} \\
$\omega_0$ & Nominal Frequency& $2\pi\times$\SI{50}{rad\per\second} \\
$L_c$ & Coupling Inductance & \SI{0.35}{\milli\henry}\\
$R_c$ & Coupling Resistance & \SI{30}{\milli\ohm} \\
$w_c$ & Filter Constant & \SI{31.4}{rad\per\second}\\
$m_p$ & Default $P-\omega$ Droop Gain & \num{9.3e-5} \si{rad\per\second\per\watt}\\
$n_q$ & Default $Q-V$  Droop Gain & \num{1.3e-3} \si{\volt\per VAR}\\
$L_{l}$ & Line Inductance & \SI{0.26}{\milli\henry\per\kilo\meter}\\
$R_{l}$ & Line Resistance & \SI{165}{\milli\ohm\per\kilo\meter}\\
$l_{ij}$ & Line Length & [5, 4.1, 3, 6] \si{km}\\
$Z_1$ & Bus 1 Load & \num{25} \si{\ohm}\\
$Z_2$ & Bus 2 Load & \num{20} \si{\ohm}\\
$Z_3$ & Bus 3 Load & \num{20+4.72i} \si{\ohm}\\
$Z_4$ & Bus 4 Load & \num{40+12.58i} \si{\ohm}\\
$Z_5$ & Bus 5 Load & \num{18.4+0.157i} \si{\ohm}\\
$X/R$ & Average X/R Ratio & 0.6224\\
\hline
\end{tabular}
\end{center}
\end{table}

\begin{table}[t!]
%\caption{Average Percentage Errors between Proposed and Simple 3rd Order Models}
\caption{Comparison of Average Percentage Errors}
\label{T:criticalkp}
\begin{center}
\begin{tabular}{c|c|c||c|c|c}
\textbf{Full} & \textbf{Proposed} & \textbf{Simple} &\textbf{Full} & \textbf{Proposed} & \textbf{Simple} \\
\hline
\multicolumn{3}{c||}{$X/R=0.622$} &\multicolumn{3}{c}{$X/R=0.913$} \\
\hline
0.64\% & 0.68\% & 1.46\% &0.78\% & 0.89\% & 2.9\% \\
\hline
\multicolumn{3}{c||}{$X/R=1.187$} &\multicolumn{3}{c}{$X/R=1.863$} \\
\hline
0.93\% & 1.12\% & 5.7\% & 1.27\% & 1.72\% & 17.39\%\\
\hline
\end{tabular}
\end{center}
\end{table}

\begin{table}[t!]
%\caption{Computation Time for EMM and ROM1}
\caption{Computational Time Comparison}
\label{T:Computation}
\begin{center}
\begin{tabular}{c||c|c||c|c}
 & \multicolumn{2}{c||}{\textbf{n = 5}} & \multicolumn{2}{c}{\textbf{n = 25}}\\
\hline
& \textbf{Full} & \textbf{Proposed} & \textbf{Full} & \textbf{Proposed}\\
\hline
$n_s$ & 42 & 15 & 222 & 75\\
\hline
ode23 & NA & 0.118s & NA & 0.119s\\
\hline
ode23s & 17.36s & 0.367s & $>$20s & 1.727s\\
\hline
ode23t &0.345s & 0.067s & 0.926s & 0.08s\\
\hline
ode23tb & 0.384s & 0.073s & 1.14s & 0.097s\\
\hline
\end{tabular}
\end{center}
\end{table}

\bibliographystyle{IEEEtran}
\bibliography{mainbib}

\begin{thebibliography}{10}
\providecommand{\url}[1]{#1}
\csname url@samestyle\endcsname
\providecommand{\newblock}{\relax}
\providecommand{\bibinfo}[2]{#2}
\providecommand{\BIBentrySTDinterwordspacing}{\spaceskip=0pt\relax}
\providecommand{\BIBentryALTinterwordstretchfactor}{4}
\providecommand{\BIBentryALTinterwordspacing}{\spaceskip=\fontdimen2\font plus
\BIBentryALTinterwordstretchfactor\fontdimen3\font minus
  \fontdimen4\font\relax}
\providecommand{\BIBforeignlanguage}[2]{{%
\expandafter\ifx\csname l@#1\endcsname\relax
\typeout{** WARNING: IEEEtran.bst: No hyphenation pattern has been}%
\typeout{** loaded for the language `#1'. Using the pattern for}%
\typeout{** the default language instead.}%
\else
\language=\csname l@#1\endcsname
\fi
#2}}
\providecommand{\BIBdecl}{\relax}
\BIBdecl

\bibitem{hatziargyriou2007microgrids}
N.~Hatziargyriou, H.~Asano, R.~Iravani, and C.~Marnay, ``Microgrids,''
  \emph{IEEE Power Energy Mag.}, vol.~5, no.~4, pp. 78--94, 2007.

\bibitem{olivares2014trends}
D.~E. Olivares, A.~Mehrizi-Sani, A.~H. Etemadi, C.~A. Canizares, R.~Iravani,
  M.~Kazerani, A.~H. Hajimiragha, O.~Gomis-Bellmunt, M.~Saeedifard,
  R.~Palma-Behnke \emph{et~al.}, ``Trends in microgrid control,'' \emph{IEEE
  Trans. Smart Grid}, vol.~5, no.~4, pp. 1905--1919, 2014.

\bibitem{Zoka2004proc}
Y.~Zoka, H.~Sasaki, N.~Yorino, K.~Kawahara, and C.~C. Liu, ``An interaction
  problem of distributed generators installed in a microgrid,'' in \emph{IEEE
  Int. Conf. Electr. Utility Deregulation, Restruct. Power Technol.}\hskip 1em
  plus 0.5em minus 0.4em\relax IEEE, Apr. 2004, pp. 795--799.

\bibitem{Jiayi2008comprehensive}
J.~Huang, C.~Jiang, and R.~Xu, ``A review on distributed energy resources and
  microgrid,'' \emph{Renew. Sustain. Energy Rev.}, vol.~12, no.~9, pp.
  2472--2483, 2008.

\bibitem{parhizi2015state}
S.~Parhizi, H.~Lotfi, A.~Khodaei, and S.~Bahramirad, ``State of the art in
  research on microgrids: a review,'' \emph{Access, IEEE}, vol.~3, pp.
  890--925, 2015.

\bibitem{chandorkar1993control}
M.~C. Chandorkar, D.~M. Divan, and R.~Adapa, ``Control of parallel connected
  inverters in standalone ac supply systems,'' \emph{IEEE Trans. Ind. Appl.},
  vol.~29, no.~1, pp. 136--143, 1993.

\bibitem{coelho2002small}
E.~Coelho, P.~Cortizo, and P.~Garcia, ``Small-signal stability for
  parallel-connected inverters in stand-alone ac supply systems,'' \emph{IEEE
  Trans. Ind. Appl.}, vol.~38, no.~2, pp. 533--542, 2002.

\bibitem{guerrero2004wireless}
J.~M. Guerrero, L.~G. De~Vicuna, J.~Matas, M.~Castilla, and J.~Miret, ``A
  wireless controller to enhance dynamic performance of parallel inverters in
  distributed generation systems,'' \emph{IEEE Trans. Power Electron.},
  vol.~19, no.~5, pp. 1205--1213, 2004.

\bibitem{hatziargyriou2013microgrids}
N.~Hatziargyriou, \emph{Microgrids: architectures and control}.\hskip 1em plus
  0.5em minus 0.4em\relax John Wiley \& Sons, 2013.

\bibitem{pogaku2007modeling}
N.~Pogaku, M.~Prodanovi{\'c}, and T.~C. Green, ``Modeling, analysis and testing
  of autonomous operation of an inverter-based microgrid,'' \emph{IEEE Trans.
  Power Electron.}, vol.~22, no.~2, pp. 613--625, 2007.

\bibitem{iyer2010generalized}
S.~V. Iyer, M.~N. Belur, and M.~C. Chandorkar, ``A generalized computational
  method to determine stability of a multi-inverter microgrid,'' \emph{IEEE
  Trans. Power Electron.}, vol.~25, no.~9, pp. 2420--2432, 2010.

\bibitem{guo2014dynamic}
X.~Guo, Z.~Lu, B.~Wang, X.~Sun, L.~Wang, and J.~M. Guerrero, ``Dynamic
  phasors-based modeling and stability analysis of droop-controlled inverters
  for microgrid applications,'' \emph{IEEE Trans. Smart Grid}, vol.~5, no.~6,
  pp. 2980--2987, 2014.

\bibitem{mariani2015model}
V.~Mariani, F.~Vasca, J.~C. V{\'a}squez, and J.~M. Guerrero, ``Model order
  reductions for stability analysis of islanded microgrids with droop
  control,'' \emph{IEEE Trans. Ind. Electron.}, vol.~62, no.~7, pp. 4344--4354,
  2015.

\bibitem{rasheduzzaman2015reduced}
M.~Rasheduzzaman, J.~A. Mueller, and J.~W. Kimball, ``Reduced-order
  small-signal model of microgrid systems,'' \emph{IEEE Trans. Sustain.
  Energy}, vol.~6, no.~4, pp. 1292--1305, 2015.

\bibitem{mohamed2008adaptive}
Y.~A.-R.~I. Mohamed and E.~F. El-Saadany, ``Adaptive decentralized droop
  controller to preserve power sharing stability of paralleled inverters in
  distributed generation microgrids,'' \emph{IEEE Trans. Power Electron.},
  vol.~23, no.~6, pp. 2806--2816, 2008.

\bibitem{nikolakakos2016stability}
I.~P. Nikolakakos, H.~H. Zeineldin, M.~S. El-Moursi, and N.~D. Hatziargyriou,
  ``Stability evaluation of interconnected multi-inverter microgrids through
  critical clusters,'' \emph{IEEE Trans. Power Syst.}, vol.~31, no.~4, pp.
  3060--3072, 2016.

\bibitem{machowski2011power}
J.~Machowski, J.~Bialek, and J.~Bumby, \emph{Power system dynamics: stability
  and control}.\hskip 1em plus 0.5em minus 0.4em\relax John Wiley \& Sons,
  2011.

\end{thebibliography}

\end{document}